\documentclass[letterpaper,10pt]{article}
\title{Networks of Moore Machines}
\author{Victor Yodaiken
}

\usepackage{arxiv}
\usepackage{arxiv}
\usepackage[T1]{fontenc}    
\usepackage{graphicx}
\usepackage{caption}
\usepackage{subcaption}
\usepackage{url}
\usepackage{hyperref}
\usepackage{amsfonts}       
\usepackage{amsmath}       
\usepackage{amsthm}       
\newcommand{\ess}{\epsilon}
\newcommand{\xy}{\cdot}
\newcommand{\concat}{\mathop{\mathtt{ concat }}}
\newtheorem{defn}{Definition}
\begin{document}


\begin{abstract}
\end{abstract}

\maketitle

\section{Introduction}
\begin{figure}[ht!]
\begin{center}\fbox{\includegraphics[width=0.4\textwidth]{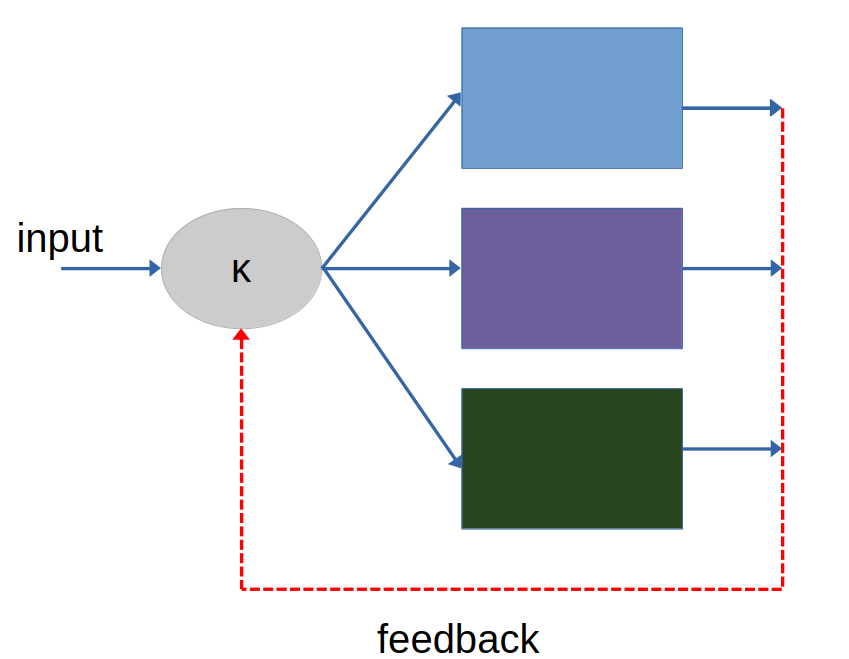}}\end{center}
\caption{The concurrent product of Moore machines\label{fig:product}}
\end{figure}

Two underappreciated ideas from the 1950s and 1960s can 
be used together to specify and understand
large scale networks of interconnected Moore machines. 
Moore machines are state machines where each state
is associated with an output value\cite{moore,hopcroft}.
The first idea is the ``concurrent product'' from an  
influential 1962 paper by Juris Hartmanis\cite{hartmanis} which 
connects and combines Moore machines in arbitrary networks (see figure
\ref{fig:product}).
The main 
topic of Hartmanis's paper was 
the ``loop-free'' or ``cascade''
restriction of the concurrent product
where information can only flow 
linearly between factors and there is no interaction, no feedback.
Loop-free products are 
related to semigroup theory and the Krohn-Rhodes theorem\cite{holcombe} (and seesection \ref{sec:monoids}). 
Systems where the flow of
information and control between factors can be bidirectional,
require feedback loops in the connection graph.

The
second idea used here 
 is an extension of primitive recursion to finite sequences
from 
R\'osza P\'eter's book on recursive functions, originally
published in 1950\cite{peter}. P\'eter described this extension in a
short {example} in the last chapter of her book
(and later in more detail\cite{petercomputer}).
P\'eter did not apply recursive
functions to automata but they can be used to represent Moore machines,
abstract properties of 
Moore machines, and products of Moore machines
without enumeration of states\cite{yodaikenlarge}. These functions, 
which are called are called
\emph{sequential functions} in this paper, are a class of maps
\[f:A^*\to Y\]
where \(A\) is a set (or alphabet) of input symbols, \(Y\) is a set of output
symbols and \(A^*\) is the set of finite sequences over \(A\).
The intuitive meaning is that for \(w\in A^*\) \(f(w)\) is the output of the 
state machine described by \(f\) in the state reached by following \(f\) from 
the initial state. Sequential function composition can mirror state machine
 products. In particular, a type of simultaneous recursion corresponds
to the concurrent product as shown in section \ref{sec:products}.

The topic of \cite{yodaikenlarge} was application
of sequential functions to specification 
and verification of computer hardware and 
software, but it's possible there may be other interesting 
applications both within and outside of
computer science. The goal of this paper 
is to make the straightforward mathematics as clear as
possible without requiring the reader to plow through e.g. detailed
explanations of distributed consensus algorithms.

Section \ref{sec:pr} reviews Moore machines, defines sequential functions,
 and establishes the relationship between the two.
Section \ref{sec:products} is mostly concerned with how
simultaneous primitive recursion defines networks (and automata products).
Section\ref{sec:bg} provides some background.

\section{Sequential functions \label{sec:pr}}

\subsection{Moore machine tuples}
Start by reviewing the definition of a Moore machine tuple. 
Moore machine tuples are not the only possible 
representations of Moore machines.
Traditional alternatives are state diagrams and transition tables.

\begin{defn} \textbf{A Moore machine tuple} is 
 \(M=(A,Y,S,s_0,\delta,\lambda)\)
where \(A\) is a set of events (or inputs)
, \(Y\) is a set of outputs, \(S\) is the 
state set, \(s_0\in S\) is the initial or start state,
 \(\delta:S\times A\to S\) is the transition map 
and \(\lambda:S\to Y\) is the output map.
If the state set is finite, say the tuple is finite state. 
\end{defn}
A key aspect of Moore machines is that the state transition function is 
encapsulated within the machine by the output map -- there is a distinction 
between interior state and visible state (output).

\subsection{Sequence preliminaries}
Let \(A^*\) be the set of finite sequences over \(A\) and 
\(w\xy a\) be the sequence obtained from \(w\in A^*\) 
by appending \(a\) to the right. The empty sequence
\(\ess\in A^*\) acts as a zero and \(w\xy a\) is like
successor for primitive recursive functions on sequences.
Let \(wz\) and \(w\concat z\) denote concatenating \(z\) to \(w\),

\subsection{Sequential functions}
\begin{defn}
\(f:A^*\to Y\) is a \textbf{sequential function} from
	\((c,h,g)\) for constant \(c\) and maps
	\(h\) and \(g\), if and only if for all \(w\in A^*\)
\[\begin{array}{l}
f(w) = h(f'(w)) \\
\\
f'(\ess) = c \text{ and } f'(w\xy a) = g(f'(w),a)
\end{array}\]
\end{defn}

In general, sequential functions may be partial.

The function \(f'\) is an algorithm for computing the state
set from the event sequence and it is encapsulated by the function \(h\).
Sequential functions are very similar to Moore machine tuples, but
the state set is implicit and functions are more convenient for some
purposes.

The behavior of one sequential function can be 
defined or constrained by referring to the behavior of other sequential
functions. If 
\(temp:Samples^*\to Y\) computes the current temperature from events which
may simply be samples like those produced by
 analog-to-digital sensors, and \(valve:Samples^*\to \{open,close\}\)
 finds the position of a valve in the same sample data, then 
 \[\text{if }temp(w) > 99C,  \text{ then }valve(w)=open\]
 is a reassuring property. If \(Status(w)\in \{running, waiting\}\)
 and we want to make sure that the system is not waiting forever,
 define
\[\begin{array}{l}Delayed(\ess) = 0\\
Delayed(w\xy a) = \begin{cases} 1+Delayed(w)&\text{if }Status(w\xy a)=waiting\\
 0&\text{otherwise}\end{cases}.\end{array}\]
Then require there is an \(n\) so that 
\(Delayed(w) < n\). It's easy to show that
\(Delayed\) is a sequential function if \(Status\) is one. 

If \(f_1:A^*\to Y_1\) is sequential,
then \( f_2(w) = H(f_1(w))\) is also sequential. If 
\(f_i:A^* \to Y_i\) for \(i=1,\dots n\), then
\(f(w) = (f_1(w)\dots ,f_n(w))\) is also sequential.
These
can be shown from the definition of sequential functions.
Simultaneous recursion, discussed below, can mimic the concurrent
product. Nevertheless, sequential functions and
Moore machine tuples represent the same mathematical objects (for
our purposes). 

\subsection{Sequential functions and Moore machine tuples}
\begin{defn}\label{defn:characteristic} The \textbf{characteristic function} of a 
Moore machine tuple  \(M=(A,Y,S,s_0,\delta,\lambda)\)
is the sequential function \(M^*: A^*\to Y\) from
\(s_0\) (the constant) and \(\lambda\) and \(\delta\):
\[\begin{array}{l}
M^*(w) = \lambda(\delta^*(w))\\
\delta^*(\ess)= s_0 \text{ and } \delta^*(w\xy a)= \delta(\delta^*(w),a)
\end{array}
\]
\end{defn}

So every Moore machine tuple defines a sequential function. 
By construction, if \(f:A^*\to Y\)
is not sequential there is no \(M\) so that \(M^* = f\). An example
of a non-sequential map \(f:A^*\to Y\) could have some \(w\) and 
\(u\) so that \(f(wu)\) is defined, 
but \(f(w)\) is not defined. That doesn't make 
physical sense for discrete state systems and it would mean that
proofs by induction on sequences would fail. 

Conversely, every sequential function
\(f:A^*\to Y\) defines a Moore machine tuple.

\begin{defn}
If \(f\) is sequential from \((c,h,g)\), let
\(S_f=\{f'(w): w\in A^*\}\), \(s_0 = f'(\ess)=c\), and \(\delta(s,a) = f'(w\xy a)\).
Then \[M_f = (A,\{h(s):s\in S_f\},S_f,c,\delta_f,h)\]
is the corresponding Moore machine tuple. Say \(f\) is \emph{finite state}
if \(S_f\) is finite.
\end{defn}
 
Alternatively, if \(f\) is sequential, the standard
equivalence
\[ w\sim_f u \text{ iff }\forall z\in A^*, f(wz) = f(uz)\]
defines a state set via the equivalence classes.

There are an infinite number of Moore machine tuples
that have the same
characteristic function as any sequential function, 
but their differences (e.g. unreachable states or
differences in the labels of state sets) are artifacts of the Moore
machine tuple representation and are not important unless we want to 
build transition tables or circuits implementing the automata. This
is the key difference in perspective between \cite{hartmanis} and the
work here and in \cite{yodaikenlarge} -- because here we are 
concerned with state systems
where it is often infeasible to construct the transition tables.  

\section{Products \label{sec:products}}

As with the arithmetic primitive recursive functions, sequential functions 
have a number of apparent extensions that do not lead out of the class.
The simplest example is just output composition as noted above.
Less simply, the function 
\[f(w) = (f_1(w),\dots f_n(w))\]
must be sequential if each \(f_i:A^*\to Y_i\) is sequential, and must be finite state if each
\(f_i\) is finite state.
This composition corresponds to the direct product of Moore
machine tuples. Conceptually, each component changes state in parallel, but
the components do not interact. Each \(f_i\) must be a function of
sequences over the  same event set, \(A^*\), but each 
can have distinct output sets. 
\begin{figure}[ht!]
\begin{center}\fbox{\includegraphics[width=0.3\textwidth]{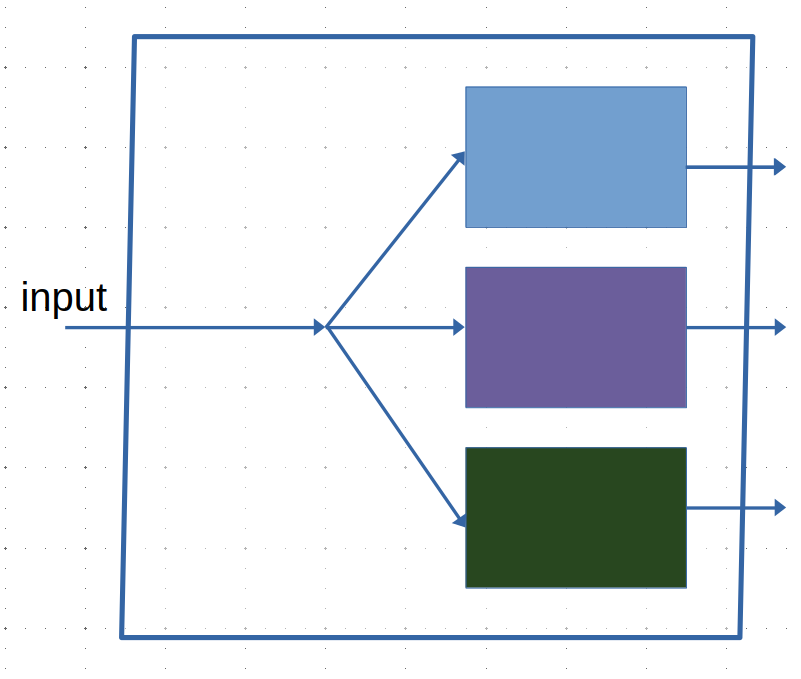}}\end{center}
\caption{The direct product of Moore machines\label{fig:direct}}
\end{figure}

If \(f(w) = (f_1(w),\dots f_n(w))\) is sequential, then \(H(f(w))\) is sequential.

\subsection{Introduction to concurrent composition}
To understand the composition that corresponds to the concurrent product,
start with the composition of two sequential functions
\(f_i:A_i^*\to Y_i\), \(i=1,2 \), to get a sense of what is involved.
The composition is partly defined by
\[f(w) = (f_1(u_1(w)),f_2(u_2(w)))\]
so  \(f(w)\in Y_1\times Y_2\) is pair valued,
Then the \(u_i\) are defined simultaneously. For this
example, the components should
start in their initial state:
\[u_i(\ess) = \ess\]
and so the value of \(f_w(\ess)\) is well defined:
\[
\begin{array}{ll}f(\ess)& = (f_1(u_1(\ess)),f_2(u_2(\ess)))\\
&= (f_1(\ess),f_2(\ess))
\end{array}\]

Suppose there is an event set \(A\) for 
the composite function. 
When the event sequence for the composite function 
 has an \(a\in A\) appended to it, the two 
components each advance by \(\kappa_i(y_1,y_2,a)\) which, in this example,
is a single event \(\kappa_i(y_1,y_2,a)\in A_i\). Putting it together:
\[\begin{array}{l}
	f(w) = (f_1(u_1(w)),f_2(u_2(w)))\\ \\
	u_i(\ess) = \ess\\
	u_1(w\xy a) = u_1(w)\xy \kappa_1(f(w),a)\\
	u_2(w\xy a) = u_2(w)\xy \kappa_2(f(w),a)\\
\end{array}\]
This composition is a type of \emph{simultaneous composition} although
\(f\) and the \(u_i\) are defined not quite simultaneously
but recursively in terms of each other. For comparison, Hartmanis's 
original concurrent product is give in section \ref{sec:hart}. 

In the more general form of the composition, there can be any number of 
components and \(\kappa_i\) is sequence valued so that the components can
advance by 0 or more steps on each step of the composite system.

\subsection{Concurrent composition}
In general the input sets and output sets of the 
functions being composed 
can be the same or distinct - the labels
of these sets do not have any magical properties. There is 
an event set \(A\) for the product which does not have to have
any intrinsic relationship to the event sets of the components.
Recall that sequential functions are determined by the triple 
\((c,h,g)\). The concurrent composition is determined by 
\((v_1,\dots v_n,\kappa)\) where each \(v_i\in A_i^*\) is the initial 
sequence for component \(i\) and \(\kappa\) is the connection function.
In the initial state of the system, each component \(i\) is started
in the state determined by \(v_i\) which could be \(\ess\) if we want
the component to begin in its initial state. The function \(\kappa\)
determines what happens to each 
component when the product system advances and depends on the values of
\(f_i(w)\) not \(f'_i(w)\). If 
\(\kappa(y_1\dots y_n,a)=(z_1\dots z_n)\), let 
\(\kappa_i(y_1\dots y_n,a)= y_i\) as a convenience.
Then:

\begin{defn}The concurrent composition of \(f_i:A_i^*\to Y_i\), \(i=1,\dots n\),
	from \(v_i\in A_i^*,\ i=1,\dots n\),
	and some  \[\kappa:Y_1\dots \times Y_n\times A\to A_1^*\dots \times A_n^*\] 
\[\begin{array}{l}f(w) = (f_1(u_1(w)),\dots f_n(u_n(w)))\\ \\
u_i(\ess)=v_i, \ u_i(w\xy a) = u_i(w)\concat \kappa_i(f(w),a)\end{array}
\]
\end{defn}

For example,  suppose we have a mesh of processing elements 
\(E_i:A^*\to Y\), \(1,\dots n\) and 
\[Mesh(w)= (E_1(u_1(w)),\dots E_n(u_n(w)))\].
Suppose each 
element \(i\) is connected to some collection of
neighbors with indexes given by
\(\mu(i)\).  Write
\(\mu(i,j)\) for the \(j^{th}\) neighbor of element \(i\) and 
\(|\mu(i)|\) for the number of neighbors of \(i\). 
If \(\bar{x}=Mesh(w)\) let \(\bar{x}:\mu(i,j)\) be the \(\mu(i,j)^{th}\)
element of \(\bar{x}\) -- the value of \(E_{\mu(i,j)}(u_{\mu(i,j)}(w))\)
if the reader can excuse this profusion of subscripts. The idea is that the set
of pairs \((m,y)\) of indexes and matching output values
of the neighbors will be assembled together as an input to each element
\[\alpha(i,\bar{x})=  \{(\mu(i,j),\bar{x}:\mu(i,j)): 0<j \leq |\mu(i)|\}\]   
These sets are the inputs for the components. We don't specify here what the 
components do with these inputs, but they receive the outputs of all their
neighbors on each step and compute some output possibly also depending
on the state of \(E_i\) (which can depend on prior inputs). 
Then let:
\[\kappa_i(\bar{X},a) = \alpha(i,Mesh(w))\]
The input \(a\) is just ignored for now but could be used to add inputs for
edge elements.

\section{Related work and background \label{sec:bg}}
Arbib \cite{Arbib} (1968) writes
\begin{quote}\emph{ ``we may say automata theory is the study of partial functions \(F:A^*\to Y^*\)''} -- (p. 6).
\end{quote}
Similarly, Pin\cite{pin} (p. 100) defines a sequential function \(\sigma:A^*\to Y^*\) as 
``a function whose behavior is defined by a machine called a `sequential transducer' ''. 
Both of these can be obtained from the characteristic map.
The set names on both of these are changed for consistency here.

Variations
of these maps
can be found all over the automata theory literature (e.g. \cite{assmus,ginzburg,holcombe,pin}).
In \cite{assmus} (p. 17) they are called input-output functions.

These
maps are primitive recursive in the sequence but either
automata theory researchers did not recognize these maps
as primitive recursive or they
didn't consider it important enough to mention. P\'eter's foundational text
on recursive functions briefly discusses primitive recursion on sequences
in an example in the last section\cite{peter} (p. 285-286) 
and motivates the example by citing  
``computing automata'', but does not explore the matter further. That
example is where
this author learned about primitive recursive functions on sequences.
A later work\cite{petercomputer}
is more expansive but does not mention automata and is more concerned
with formal languages and computability. 
See also Eilenberg and Elgot\cite{eilenberg}.

Moore's 1954 paper\cite{moore} defined what are now
called Moore machines although they are not too different from 
Huffman's sequential circuits\cite{huffman} (which is the inspiration for
the name \emph{sequential functions}). 

\subsection{The state machine product}\label{sec:hart}
Hartmanis's definition of the concurrent product\cite{hartmanis} is
quoted below. The main difference with the version used here is that in
the version used here components advance by a sequence (which may be empty).
This is easy to do for the function representation and somewhat
more cumbersome
for the Moore machine tuple version. 
\begin{quote} \emph{
DEFINITION 1. Let \(M_1, M_2, ., M_n,\) be a set of Moore type machines
	in which the outputs of any machine \(M_i, i = 1, 2,..., n\),
may be used as inputs to other machines. We shall say that this set
of interconnected machines is concurrently operating if the next
state (state at time t + 1) of each machine \(M_i\) depends on the present
state of \(M_i\), the present outputs (which depend only on the present
states) of the machines to which it is connected, and the present
external input.
The ordered n-tuple (or "configuration") of the present states of
the machines \(M_1, M_2,. , M_n\) will be referred to as the state of the
	interconnected machine.} -- Hartmanis\cite{hartmanis}, p 117.
\end{quote}

At each step, each component \(M_i\) gets input
that is a function of the external input (the input to the 
system) and the outputs of some or all of the components.
To illustrate suppose, for \(0< i\leq n\), 
\(M_i=(A_i,Y_i,S_i,s_{i,0},\delta_i,\lambda_i)\) and we want to connect
them as factors of a product state machine
\[M= (A,Y,S,(s_{1,0}\dots,s_{n,0}),\delta,\lambda).\]
Let \(A\) be the product input alphabet  (the inputs
for the system), \(Y= Y_1\dots \times Y_n\) is the output set of the product
machine and \(S= S_1\dots \times S_n\) is the state set of the product machine.
For \(s = (s_1,\dots s_n)\) let
\[\lambda(s) = (\lambda_1(s_1)\dots , \lambda_n(s_n)).\]
To complete the product, we need
a connection map:
\[\gamma:Y\times A\to A_1\dots \times A_n\] 
When the product machine is in state \(s\) an input \(a\) will advance \(M_i\)
by \(\gamma_i(\lambda(s),a)\), which is some element of \(A_i\).
That is, if \(\gamma(y,a) = (y_1,\dots ,y_n)\) then \(\gamma_i(y,a)=y_i\).
The input alphabets
of the component machines may be distinct or not, depending on what is being 
modelled. 
The product transition map \(\delta\) is given by:
\[\delta(s,a) = (\delta_1(s_1,\gamma_1(\lambda(s),a))\dots
,\delta_n(s_n,\gamma_n(\lambda(s),a)))\]

Hartmanis and Stearns defined a slightly different version of the concurrent
 product 
they called a \emph{network} in their book\cite{HartmanisStearns}.
Gecseg\cite{Gecseg}
has another version and 
cites other work dating back to the 1970s and earlier.
Interconnected
 products seem to have been considered obvious in classical 
automata theory. See
for example a passing mention in Assmus and Florentin \cite{assmus} (p. 30) and
\begin{quote}\emph{Single output automata with given time delays can be combined into a new automaton} -- Von Neumann \cite{vonneumanninshannon} (p.45). 
\end{quote}

Both Hartmanis and
Gecseg show how by limiting feedback, the interconnection can be varied.
The connector
map 
determines the connection graph.
In the cascade  product the factor automata are ordered so that
inputs generated for \(M_i\) depend only on the outputs
of \(M_1,\dots M_i\)
and ignore the outputs of the remaining factors. 

\subsection{Semigroups}\label{sec:monoids}

The algebraic automata theory literature which 
includes Hartmanis and many others\cite{ginzburg,Arbib, holcombe,pin} 
was focused on the cascade product because it preserves the structure 
of the automata and the underlying semigroup structure.

The relationship between state machines, state machines products, and factoring
semigroups and monoids is extensively discussed in algebraic automata
theory. See \cite{holcombe} for a survey.

There are several similar
 ways to construct a monoid from a state machine. One way is to use
the characteristic sequential functions of definition \ref{defn:characteristic} and define a congruence.

\[\begin{array}{l}
\text{For }w,u\in A^* \text{ define } (w = u \bmod M) \text{ if and only if }\\
\quad (\forall z_1,z_2\in A^*)(M^*(z_1\concat w\concat z_2) = 
M^*(z_1\concat u\concat z_2))\end{array} \] 

Then the elements of the monoid are the equivalence classes 
\([w]_M = \{u\in A^*: u \sim w \bmod M\}\). The monoid
operation is \([w]_M\circ [u]_M = [w\concat u]_M\) and the identity element
is \([\ess]_M\).

If a monoid has a cancelation property, it is a group. An automaton with 
a monoid that is a group is called a group machine. 
Krohn-Rhodes theorem \cite{krohnrhodes} surprisingly ties factoring 
of finite automata using loop-free (cascade)
 products to the Jordan-Holder theorem of group theory\cite{maclane} (p. 430-432). As summarized
by Zeiger\cite{zeigerchapter}, one consequence of Krohn-Rhodes is:
\begin{quote}\emph{ Each finite state automaton can be built as a cascade of two
state automata and simple-group automata}
(p. 77)\end{quote} 

The general product of section \ref{sec:products} does not preserve the 
group structure of automata. In fact, if \(M\) has \(n\) states, it can
be factored to a product of \(\lceil \log_2 n \rceil\) two state automata.
In the general case, each transition of the product
transmits the one bit state of every factor to every other factor  and
the state of the original automaton is reconstructed from the bit sequence.
These factorizations can be limited either by limiting the
connectivity of the connection graph or by limiting the quantity of information carried by the connection map (the bandwidth) but there doesn't seem to be
much work on either beyond what was known in the 1960s. See Gecseg \cite{Gecseg}
 for a survey.

One striking aspect of the older algebraic automata theory
literature is that the problem that this work
is trying to solve -- how to characterize systems with too many states to
enumerate -- was much less interesting. Researchers in classical automata
theory seem to have generally assumed that writing down the state table
of a state machine was not a problem. The problem was factoring and
minimizing that state machine. This difference 
is a reflection of the change in circuit technology scale.

\bibliographystyle{alphaurl}
\bibliography{cellular.bib}
\end{document}